\documentclass[12pt]{iopart}
\usepackage{textcomp}
\usepackage{graphicx}
\usepackage[english]{babel}
\usepackage{float}

\usepackage[bf, hang]{subfigure}
\begin{document}
\begin{center}
\title[Monte Carlo Simulations of the LARES space
 experiment]{Monte Carlo Simulations of the LARES space
 experiment to test General Relativity and fundamental physics}
\end{center}

\author{I Ciufolini$^1$, B Moreno Monge$^2$, A Paolozzi$^3$, R Koenig$^2$, G Sindoni$^3$,  G Michalak$^2$ and E Pavlis$^4$}

\address{  $^1$ Dipartimento di Ingegneria dell'Innovazione, Universit\`a del Salento, Lecce, and Centro Fermi, Roma  (Italy)}
\address{ $^2$ GFZ German Research Center for Geosciences (Germany)}
\address{ $^3$ Scuola di Ingegneria Aerospaziale and DIAEE, Sapienza Universit\`a di Roma (Italy)}
\address{ $^4$ Goddard Earth Science and Technology Center, University of Maryland, Baltimore County (USA)}
\ead{ ignazio.ciufolini@unisalento.it}
\begin{abstract}
The LARES (LAser RElativity Satellite) satellite was successfully launched in February 2012.
The LARES space experiment is based on the orbital determinations of the laser ranged satellites LARES,
LAGEOS (LAser GEOdynamics Satellite) and LAGEOS 2 together with the determination of the Earth's gravity field
 by the GRACE (Gravity Recovery And Climate Experiment) mission. It will test some fundamental physics
predictions and provide accurate measurements of the frame-dragging effect predicted by Einstein's theory
of General Relativity. By one hundred Monte Carlo simulations of the LARES experiment,
with simulations of the orbits of LARES, LAGEOS and LAGEOS 2 according to the latest GRACE gravity
field determinations, we found that the systematic errors in the measurement of frame-dragging amount to about 1.4\% of
the general relativistic effect, confirming previous error analyses.

\end{abstract}

\maketitle

\section{The LARES space experiment to test General Relativity and fundamental physics}
LARES (LAser RElativity Satellite) is a laser ranged satellite of the Italian Space Agency (ASI) see Fig. 1.
It was launched successfully on February 13, 2012 with the qualification flight of VEGA,
the new launch vehicle of the European Space Agency (ESA), which was developed by ELV (Avio-ASI)$^{1,2}$. 
Laser ranging is the most accurate technique for measuring distances to the Moon$^{3}$ and to artificial
and laser ranged satellites$^{4}$. Short-duration laser pulses are emitted from lasers on Earth and
 then reflected back to the emitting laser-ranging stations by retro-reflectors on the Moon or on
 artificial satellites. By measuring the total round-trip travel time we are now able to determine
the instantaneous distance of a retro-reflector on a laser ranged satellite with precision of a
 few millimetres$^{5}$. Laser ranging activities are organized under the International Laser Ranging
Service (ILRS)$^{3}$ which provides global satellite laser ranging (SLR) and lunar
 laser ranging data and their derived products to support geodetic, geophysical,
and fundamental physics research activities.

\begin{figure}[H]
\label{fig1}
\begin{center}
\includegraphics[width=0.8\textwidth]{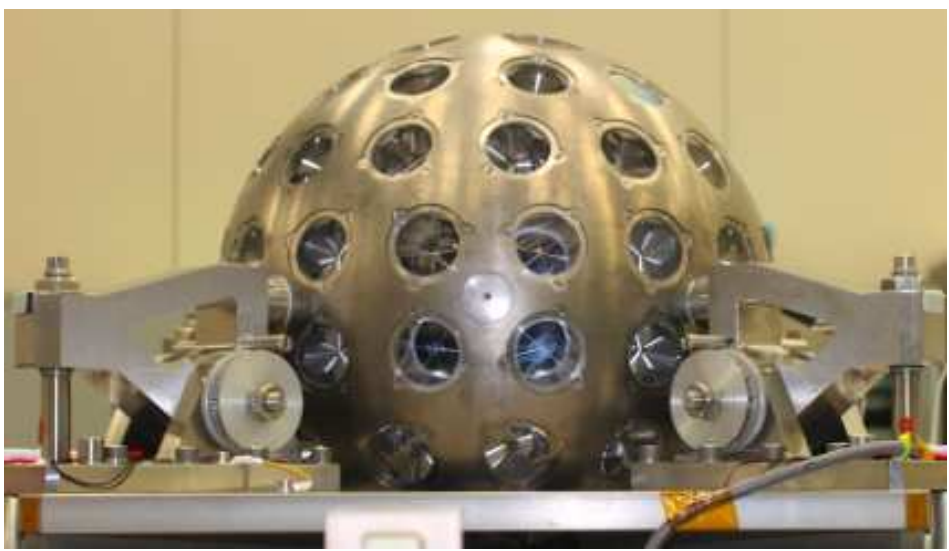}
\caption{The LARES satellite on the separation mechanism before launch (Courtesy of ASI).}

\end{center}
\end{figure}

LARES has the highest mean density of any known object orbiting in the Solar System.
It is spherical and covered with 92 retro-reflectors, and it has a radius of 18.2 cm.
 It is made of a tungsten alloy, with a total mass of 386.8 kg, resulting in a ratio of cross-sectional
area to mass that is about 2.6 times smaller than that of the two LAGEOS (LAser GEOdynamics Satellite)
satellites. Before LARES, the LAGEOS satellites had the smallest ratio of cross-sectional area to mass
 of any artificial satellite$^{4}$. LAGEOS and LAGEOS 2 have an essentially identical structure but
 different orbits. They are passive, spherical satellites covered with retro-reflectors and made
 of heavy brass and aluminium alloy. The mass of LAGEOS is 407 kg and that of LAGEOS 2 is 405.4 kg;
their radius is 30 cm. Their ratio of cross-sectional area to mass is approximately 0.0007 $m^2/kg$.
LAGEOS was launched by the National Aeronautics and Space Administration (NASA) in 1976 and LAGEOS 2
by NASA and the Italian Space Agency (ASI) in 1992. The semi-major axis of the LAGEOS orbit is
approximately 12,270 km, the period is 3.76 h, the eccentricity is 0.004 and the inclination is $109.8^\circ$.
 LAGEOS 2 is at a similar semi-major axis of about 12,160 km, but its orbital eccentricity is larger (0.014)
 and it has a lower inclination ($52.6^\circ$).

The LARES orbital elements are as follows: the semi-major axis is 7820 km, orbital eccentricity 0.0007,
and orbital inclination $69.5^\circ$. It is currently well observed by the global ILRS station network.
The extremely small cross-sectional area to mass ratio of LARES, i.e. 0.00027, which is smaller
than that of any other artificial satellite, and its special structure, a solid sphere with high
thermal conductivity, ensure that the unmodelled non-gravitational orbital perturbations are smaller
 than for any other satellite, in spite of its lower altitude compared to LAGEOS. This behaviour
has been confirmed experimentally using the first few months of laser ranging observations$^{6}$.

The LARES satellite, together with the LAGEOS and LAGEOS 2 satellites and the GRACE
(Gravity Recovery And Climate Experiment) mission Earth gravity field determinations$^{7,8}$,
 will accurately measure the frame-dragging effect predicted by General Relativity$^{9,10,11}$.

\subsection{Frame-dragging, General Relativity and String Theory}

In Einstein's gravitational theory the local inertial frames have a key role$^{12,13,14}$.
The strong equivalence principle, at the foundations of General Relativity, states that the
gravitational field is locally “unobservable” in the freely falling frames and thus, in these
local inertial frames, all the laws of physics are the laws of Special Relativity. The axes of
these non-rotating, local, inertial frames are determined by free-falling torque-free test-gyroscopes,
i.e. sufficiently small and accurate spinning tops. In other words, the freely falling test-gyroscopes
define axes fixed relative to the local inertial frames, where the equivalence principle holds,
that is, where the gravitational field is locally “unobservable”. If we rotated with respect to
 these gyroscopes, we would feel centrifugal forces, even though we may not rotate at all with
 respect to the “distant stars”. Indeed, the local inertial frames are determined, influenced
 and dragged by the distribution and flow of mass-energy in the Universe. Therefore, the test-gyroscopes
are also dragged by the motion and rotation of nearby matter, such as a spinning mass$^{12,13,14}$,
i.e. their orientation changes with respect to the distant stars: this is the “dragging of inertial frames”
 or “frame-dragging”, as Einstein named it in a letter to Ernst Mach$^{9}$. Frame-dragging phenomena,
which are due to mass currents and mass rotation, have also been called gravitomagnetism$^{14,15}$
 because of a formal analogy of electrodynamics with General Relativity (in the weak-field and
slow-motion approximation).

Frame-dragging affects clocks, electromagnetic waves$^{16,17}$, orbiting particles$^{10}$ and gyroscopes$^{18,19}$, e.g.
the gyroscopes of the GP-B (Gravity Probe B) space experiment.
 Frame-dragging was measured using the LAGEOS and LAGEOS 2 satellites with an accuracy of about
 10\%$^{1,20,21}$ and by the GP-B space experiment with an accuracy of about 19\%$^{22}$.
 See, for example, a description of these measurements in [23].

The angular momentum of a spinning body generates a precession of the spin axis of a test-gyroscope.
Similarly, the orbital plane of a planet, Moon or satellite is a huge gyroscope that “feels”
general relativistic effects. Indeed, frame-dragging of an orbiting test-particle manifests itself
 in the secular rate of change of the longitude of the nodes of the test-particle, i.e. of its
orbital angular momentum vector, and in the secular rate of change of the longitude of the
pericentre of the test-particle, i.e. of its Runge-Lenz vector. The rate of change of the nodal
longitude,  $ \bf \Omega$, of a test-particle or satellite, known as the Lense-Thirring effect$^{10}$, is

\begin{equation}
{\dot {\bf \Omega}}_{Lense-Thirring} = {{2 G {\bf {J}}} \over {c^2 a^3 (1-e^2)^{3/2}}}
\end{equation}

where  $ \bf \Omega$ is the longitude of the nodal line of the satellite
(the intersection of the satellite orbital plane with the equatorial plane of the central body),
 \textbf{J} is the angular momentum of the central body, \textit{a} is the semimajor axis of the orbiting test-particle,
\textit{e} is its orbital eccentricity, \textit{G} is the gravitational constant, and \textit{c} is the speed of light.

In 2008, Smith et al. showed that String Theories of the type of Chern-Simons gravity, i.e.
with action containing invariants built with the Riemann tensor ‘squared’, predict an additional
 drift of the nodes of a satellite orbiting a spinning body$^{24}$. Then, using the frame-dragging
measurement obtained with the LAGEOS satellites, Smith et al. set limits on such String Theories that may be related to dark energy and quintessence. In particular, they have set a lower
limit on the Chern-Simons mass that is related to more fundamental parameters, such as the
 time variation of a scalar field entering the Chern-Simons action, possibly related to a quintessence field. LARES will improve this limit on such fundamental physics theories.
In summary, LARES will not only accurately test and measure frame-dragging, a basic prediction of Einstein's theory of General Relativity, but it will also provide a test of some predictions of fundamental physics theories of unified interactions possibly related to dark-energy.

\subsection{How to measure frame-dragging with LARES, LAGEOS and LAGEOS 2 }

SLR can measure the positions of LARES and LAGEOS with an uncertainty of a few millimetres and their nodal longitude with an uncertainty of a fraction of a milliarcsec per year$^{5,21,29}$.
The Lense-Thirring drag of the orbital planes of the LARES and LAGEOS satellites, Eq. (1), is approximately 31 milliarcseconds per year for LAGEOS and LAGEOS 2 and 118
 milliarcseconds per year for LARES$^{1,21,25-29}$, corresponding to approximately 1.85 m/yr at the LAGEOS altitude and 4.5 m/yr at the LARES altitude. Thus the Lense-Thirring effect can
 be very accurately measured on the LAGEOS and LARES orbits if all their orbital perturbations can be modelled accurately enough$^{1,21,25-29}$.

The orbital perturbations of a satellite can be considered of gravitational and non-gravitational origin. For the LARES and LAGEOS satellites the uncertainties in the modelling of the orbital perturbations of gravitational
 origin are mainly due to the uncertainties in the Newtonian gravitational field and are by far larger than the uncertainties of non-gravitational origin, i.e. radiation pressure, both from Sun and Earth, thermal thrust
 and particle drag$^{1,26,29}$. Indeed, the LAGEOS satellites, and especially the LARES, satellite are extremely dense spherical satellites. In particular, the extremely small cross-sectional area to mass ratio of
 LARES, smaller than that of any other artificial satellite, and its special structure, a solid sphere with high thermal conductivity$^{1}$, ensure that the unmodelled non-gravitational orbital perturbations
are smaller than for any other satellite. The unmodelled effect of thermal thrust and particle drag are smaller on LARES than on LAGEOS in spite of its lower altitude compared to LAGEOS. This behaviour has
 been experimentally confirmed by the first few months of laser ranging observations$^{6}$.
Among the orbital perturbations of the node of an Earth satellite of gravitational origin, by far the largest ones are due to the Earth's deviations from spherical symmetry, both its static and time-dependent
 part$^{26,29,31}$. In particular, the only secular gravitational perturbations affecting the orbit of an Earth satellite are due to those deviations that are symmetric with respect to the Earth's spin axis
 (i.e. axially symmetric) and symmetric with respect to the Earth's equatorial plane, i.e., the so-called even zonal harmonics. The Earth's gravitational field and its gravitational potential can be
 expanded in spherical harmonics, and the even zonal harmonics are those harmonics of even degree and zero order$^{31}$. In particular, the flattening of Earth, corresponding to the second degree zonal harmonic, $\textit{J}_2$, describing the Earth's quadrupole moment, produces the largest secular perturbation of the node of an Earth satellite$^{31}$. But thanks to the observations of geopotential mapping missions such as GRACE$^{7,8}$, the Earth's shape and its gravitational field are extremely well known. 

Today, the GRACE$^{7,8}$ space mission of NASA and Deutsche Forschungsanstalt fur Luft und Raumfahrt (DLR) - the Principal Investigators being from the University of Texas Center for Space Research
 (UT CSR) and from the German Research Centre for Goesciences (GFZ) - has dramatically improved the accuracy in the determination of the Earth's spherical harmonics. For example, the dynamical flattening of the Earth's gravitational potential, $\textit{J}_2$, is today measured$^{32}$ with a relative uncertainty of only about one part in 10$^{7}$, and this uncertainty, propagated on the node of LAGEOS, corresponds in order of magnitude to the Lense-Thirring effect on its node. The GRACE spacecraft were
 launched into a polar orbit at an altitude of approximately 400 km and with a spacing between the two satellites of about 200-250 km. The spacecraft range to each other using radar and they
 are tracked by the Global Positioning System (GPS) satellites. By analyzing the GRACE Earth gravitational field determinations and their uncertainties in the even zonal harmonics, we found of
 course that by far the main source of error in the determination of the Lense-Thirring effect is due to the uncertainty in the lowest degree even zonal harmonics and in particular due to $\textit{J}_2$,
the Earth's quadrupole moment$^{20,21}$.

In order to eliminate the nodal rate uncertainties due to the errors in the Earth's even zonal harmonics, the use of two or more LAGEOS satellites was proposed$^{25,26}$. One technique proposed
the  use of two LAGEOS satellites with supplementary inclinations in order to eliminate the uncertainties due to all the even zonal harmonics$^{25}$, while another technique proposed to eliminate
the effect of the first, lowest order, N-1 Earth even zonal harmonics and to measure the Lense-Thirring effect by using the nodes of N satellites of LAGEOS type. This technique to use N observables
 to cancel the effect of the N-1 even zonal harmonic errors and to measure frame-dragging is described in [26], with explicit calculations in [27]. Following this method, by eliminating the
 nodal rate uncertainties of LAGEOS and LAGEOS 2 due to $\delta  {J}_2$, a measurement of frame-dragging with an accuracy of about 10\% was obtained, mainly due to the uncertainty in the next lowest order even zonal harmonic, ${J}_4$$^{1,11,21}$. The LARES satellite will introduce one additional observable, the node of LARES, to eliminate the uncertainty due to $\delta  {J}_4$ and to obtain an uncertainty in the measurement of frame-dragging of about 1\% only$^{28}$. 

A number of papers have been published that analyse all the error sources, of both gravitational and non-gravitational origin,
 that can affect the LARES and LAGEOS experiments$^{1,21,26-29,30,33}$. The error in the LARES experiment due to each even zonal harmonic up to degree 70 is analysed in detail in [1, 28].
The large errors due to the uncertainties in the first two even zonal harmonics, of degree 2 and 4, i.e., $\delta  {J}_2$ and $\delta  {J}_4$, are eliminated using the three observables, i.e., the three nodes of the LARES, LAGEOS and LAGEOS 2 satellites. The conclusion is that the error due to each even zonal harmonic of degree higher than 4 is considerably
less than 1\% of the Lense-Thirring effect and in particular that the error is negligible for the even zonal harmonics of degree higher than 26. This can be seen in detail in Figs. 3-6 of [28] that show the percentage errors in the measurement of the Lense-Thirring effect in the LARES experiment due to each individual uncertainty of each even zonal harmonic, corresponding to the GRACE models EIGEN-GRACE02S
 (GFZ, Potsdam, 2004) and GGM02. These figures also show th upper bound to the maximum percentage errors due to each even zonal harmonic.
This upper bound was obtained by considering as uncertainty for each harmonic the difference between the value of that harmonic in the EIGEN-GRACE02S model minus its value in the GGM02S model;
 this is a standard technique in space geodesy to estimate the reliability of the published uncertainties of comparable models. However, we stress that newer determinations of the Earth gravity field available today are much more accurate than the older EIGEN-GRACE02S and GGM02S models, therefore those previous error analyses of ours are a worst case scenario.
The purpose of the present paper is to augment the previous error analyses$^{1,28}$ by a number of Monte Carlo simulations.

\section{Design of the Monte Carlo simulations for testing the LARES experiment}
We design Monte Carlo simulations to reproduce as closely as possible the real 
experiment to measure frame-dragging using LARES, LAGEOS, LAGEOS 2 and the
 GRACE Earth gravitational field determinations. 
The simulations are performed as follows. The first step is to identify a set of physical parameters whose uncertainties have a critical impact on the accuracy of the measurement of
 the frame-dragging effect using LARES, LAGEOS and LAGEOS 2. Then, we consider the values of these critical parameters, determined either by the GRACE space mission
 (in the case of the Earth gravitational field parameters) or by previous extensive orbital analyses (in the case of the radiation pressure parameters of the satellites).
 Together with the values of these parameters, we consider their realistic 1-sigma uncertainty estimated by also taking into account the systematic errors.
The Earth gravitational field model that is considered is EIGEN-51C$^{34}$. 
EIGEN-51C is a combined global gravitational field model to degree and order 359. It consists of 6 years of CHAMP and GRACE data
 (October 2002 to September 2008) and a global gravity anomaly data set. 
So no LAGEOS data were used to derive this gravitational field model.
 The values of the critical parameters and of their 1-sigma
 uncertainties are given in Table 1. In the previous section 1.2 we have reported a brief description of the main error sources in the LARES experiment and the analysis of the errors due to the even zonal harmonic uncertainties up to degree 70$^{28}$.
\begin{table}[h!]

\caption{\label{tab2}The parameters considered in the Monte Carlo simulation with their sigmas 
(see section 2).}
\begin{indented}

\item[]\begin{tabular}{@{}lll}
\br
\textbf{Parameter} & \textbf{Nominal value}& \textbf{1-Sigma}\\
\mr
$GM$&0.3986004415E+15&8E+05\\

$C_{2, \, 0}$&-.484165112E-03&2.5E-10\\

$C_{4,\, 0}$& 0.539968941E-06&0.12280000E-11\\

$C_{6, \, 0}$&-.149966457E-06 &0.73030000E-12 \\

$C_{8, \, 0}$&0.494741644E-07&0.53590000E-12 \\

$C_{10\, , 0}$&0.533339873E-07 &0.43780000E-12 \\

$\dot{C}_{2, \, 0}$&0.116275500E-10&0.01790000E-11 \\

$\dot{C}_{4, \, 0}$&0.470000000E-11 &0.33000000E-11 \\

$C_r$ LAGEOS 1&1.13&0.00565\\

$C_r$ LAGEOS 2& 1.12 &0.0056 \\

$C_r$ LARES&   $\sim$ 1& 0.0054\\

\end{tabular}
\end{indented}
\end{table}

The value of $GM$, the geocentric gravitational constant, used herein is consistent with EIGEN-51C, i.e., is set by definition in the EIGEN-51C gravity field model development. Its value and standard deviation are taken from
the International Earth Rotation and Reference Systems Service (IERS) Conventions 2010$^{32}$, where the current knowledge of this parameter is documented.
The values of the first few even zonal harmonics of the gravity field model, i.e. $C_{2,\, 0}$ to $C_{10,\, 0}$, are also taken from EIGEN-51C, their standard deviations are calibrated values
 besides the standard deviation of $C_{2,\, 0}$ which is an estimate from comparisons of different gravity field models plus a safety margin established by multiplying the result
by a factor 3. The solar radiation coefficients, $C_r$, of the LAGEOS satellites and their standard deviations are taken from long-term use in the geodetic community,
 while those for LARES have been newly established on the basis of extensive orbital analyses of 10 months of LARES data$^{6}$.

The second step is to randomly generate 100 samples of values for each parameter of Table 1, with population distributed as a normal (Gaussian) distribution around the mean value of each parameter that is equal to its value reported in Table 1, and with standard deviation equal to the calibrated sigma of that parameter, also reported in Table 1.

 In other words, by repeating 100 times the process of the random generation of the parameters of Table 1, we obtain 100 samples of these parameters and the distribution of the values of each parameter follows a Gaussian distribution with mean and standard deviation defined in Table 1.
The third step is to generate the orbit of each satellite, i.e. LARES, LAGEOS, and LAGEOS 2. This is done by using the orbital propagator and estimator EPOS-OC$^{35}$, using each time, as input,
the values of one of the 100 sets of eleven parameters that were generated at the second step.
 The frame-dragging effect is always kept equal to its General Relativity value. These 100 simulations represent 100 approximations of the real orbit, of the three satellites, generated by
physical perturbations that are partially unknown because of the uncertainties in the parameters of Table 1. In a second orbital propagation,
we generate the orbit of each satellite, starting with the same initial conditions as the previous case but using the nominal value of each of the considered parameters and zero frame-dragging effect.
 This second set of simulations of the reference orbit of each satellite represents the set of orbits of each satellite as modelled in the orbital data analyses using the orbital
 estimators of the LARES team$^{6}$, i.e., EPOS-OC (or GEODYN or UTOPIA). Finally, for each day, we take the difference of the nodes between these two sets of orbits, i.e., between each one of the 100 cases described above and the nominal case obtained using EPOS-OC, and save that difference
 for each day and for each satellite. In this way, for a period of about three years (1097 days exactly), we obtain 100 sets of 1097 simulated residuals for each satellite,
 representing the noisy residuals of each satellite that will be obtained in the real data analysis. These 100 sets of simulated nodal residuals for each satellite provide a reasonably large number of
cases according to the Central Limit theorem.

In summary, for each of the 100 simulations we generate 1097 simulated residuals for each of the three satellites, LARES, LAGEOS and LAGEOS 2. Each of these simulations,
with the corresponding residuals, is obtained with one of the sets of physical parameters generated according to the mean and sigma reported in Table 1.
Then, the node residuals of LARES, LAGEOS and LAGEOS 2 are combined according 
to the formula (see, for example, 28):
\begin{equation}
{\delta \, \dot {\Omega}^{LAGEOS} + k \, \delta \, \dot{\Omega}^{LAGEOS 2} + h \, \delta \, \dot{\Omega}^{LARES} = (30.7 + k 31.5 + h 118)\,  milliarcsec/yr}
\end{equation}									
where \textit{k} and \textit{h} are suitable numerical coefficients chosen to eliminate the uncertainties $\delta \textit{J}_2$ and $\delta \textit{J}_4$ in the lowest even zonal harmonics
 $\textit{J}_2$ and $\textit{J}_4$, and where $30.7 \; milliarcsec/yr$,  $ 31.5 \; milliarcsec/yr$ and $118 \; milliarcsec/yr$  are the nominal values of frame-dragging predicted by General
Relativity for each satellite.
The next step, for each of the 100 simulations, is to integrate the 1097 combined residuals to get, for each day, 
the cumulative residual shift of the combination of the three nodes. 
Finally, the combination of the integrated residuals of the three satellites is fitted with a straight line using the least squares method, 
to obtain, for each of the 100 simulations, the simulated, measured value of the frame-dragging effect.
In addition (corresponding to Figures 2A, 2B and 2C), for each of the three satellites and for each of the 100 simulations, the 1097 residuals 
are integrated to get, for each day, the cumulative residual shift of the node of each of the three satellites LARES, LAGEOS and LAGEOS 2.

\section{Results of the Monte Carlo simulations}
In Figures 2 and 3 we show the results of the 100 simulations. Figure 2 shows the simulated nodal drifts corresponding to each of the 100 simulations. 
Figures 2A, 2B and 2C refer respectively to LAGEOS, LAGEOS 2 and LARES. Figure 2D refers to the LARES, LAGEOS and LAGEOS 2 combination, i.e.,
each nodal drift shown in Figure 2D (see its magnification in Figure 3B) was obtained by combining the simulated cumulative (integrated) node residuals of the three satellites for each of the
100 simulations. The simulated measured value of frame-dragging
was then obtained by fitting these raw residuals with a straight line. 
Figure 3, using a scale magnified by about 33.3 times with respect to Figure 2, shows the simulated measured nodal drifts corresponding to each of the 100 simulations. 
Figure 3A refers to the LAGEOS and LAGEOS 2 combination, whereas Figure 3B to the LARES, LAGEOS and LAGEOS 2 combination.
The result of the 100 simulations for the LARES, LAGEOS and LAGEOS 2 combination is that the mean value of the measured 
frame-dragging effect is equal to 100.24\% of the frame-dragging effect
predicted by General Relativity, with a standard deviation equal to 1.4\% of the frame-dragging effect predicted by General Relativity.

The 1.4\% uncertainty represents the {\it systematic errors} in the measurement of frame-dragging with the 
LARES experiment. In order to preliminarily evaluate the {\it statistical error} in the frame-dragging measurement, 
we have considered the standard deviation of the post-fit residuals of the preliminary, rough, orbital analyses of 
LARES, LAGEOS and LAGEOS 2 performed since the launch of LARES. In the preliminary, rough, 
fits of the combination of LARES, LAGEOS and LAGEOS 2, the standard deviation of the post-fit residuals 
is about 5 milliarcsec, i.e., approximately the same as the standard deviation of the post-fit residuals of the 
combination LAGEOS and LAGEOS 2 only (over the same period). 
This statistical error of the preliminary, rough, fits of the combination of LARES, LAGEOS and LAGEOS 2 agrees well
with the standard deviation of the post-fit residuals of the older, 2004, orbital analyses 
with LAGEOS and LAGEOS 2 only, obtained using the older model EIGEN-GRACE02S, indeed the standard deviation of the post fit 
residuals of the 2004 analyses was also approximately 5 milliarcsec. 
Since the combined 
frame-dragging effect is approximately 50 milliarcsec/yr, the statistical error over a period of 10 years would then be 
about 1\% of the total frame-dragging effect. 
Furthermore, by considering present and future improvements in the modeling of the various 
perturbations with respect to the older analyses with EIGEN-GRACE02S, due to improved sets of perturbations in 
the orbital estimators, including tides, non-gravitational perturbations, etc., the statistical error should be approximately 1\% 
of the total combined frame-dragging effect even over a shorter observational period.

\begin{figure}[H]
\label{fig2}
\begin{center}
\includegraphics[width=0.85\textwidth]{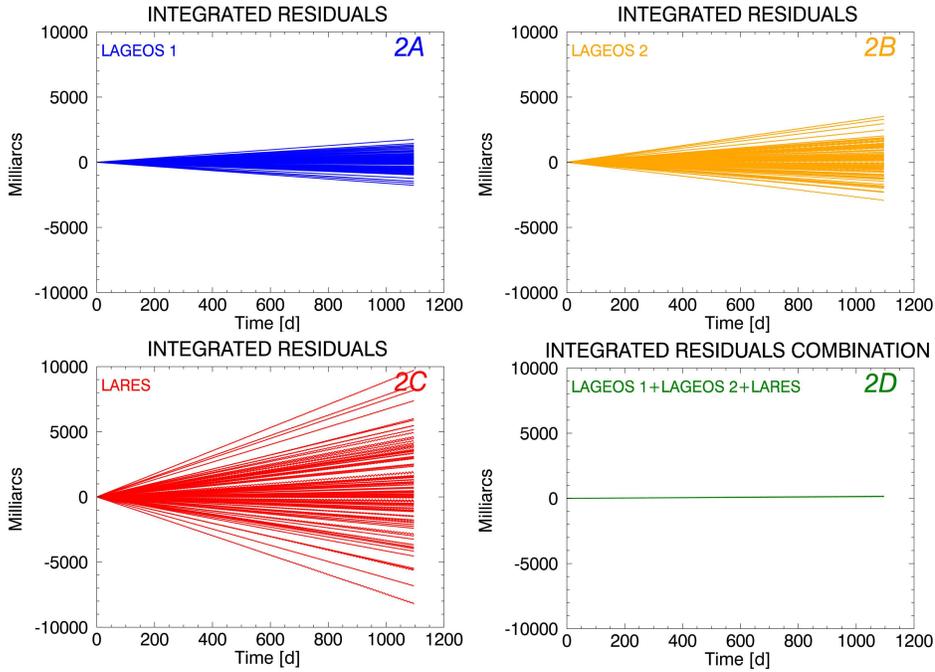}
\caption{Figures 2A, 2B and 2C show, respectively for the LAGEOS, LAGEOS 2 and LARES satellites, the simulated cumulative (integrated) residuals of the nodal longitude for each of the 100 performed simulations. Figure 2D, using the same scale, shows the simulated cumulative residuals of the combination of the nodal longitudes of LAGEOS, LAGEOS 2 and LARES, for each of the 100 simulations. That combination allows, according to formula (2), to eliminate the uncertainties in both the Earth's moments $J_2$ and $J_4$. Figure 2D clearly display the reduction of the spread between the 100 simulations, i.e., the reduction of the standard deviation of the slopes of the 100 simulations, when both the $\delta J_2$ and $\delta J_4$ uncertainties are removed from the residual nodal drifts of the satellites by using a suitable combination of their nodal residuals. The use of LARES, together with LAGEOS and LAGEOS 2, dramatically reduces the standard deviation of the slopes of the nodal residuals of the 100 simulations, that is the uncertainty in the simulated measurement of frame-dragging using LARES, LAGEOS and LAGEOS 2.}
\end{center}
\end{figure}

\begin{figure}[H]
\label{fig3}
\begin{center}
\includegraphics[width=0.85\textwidth]{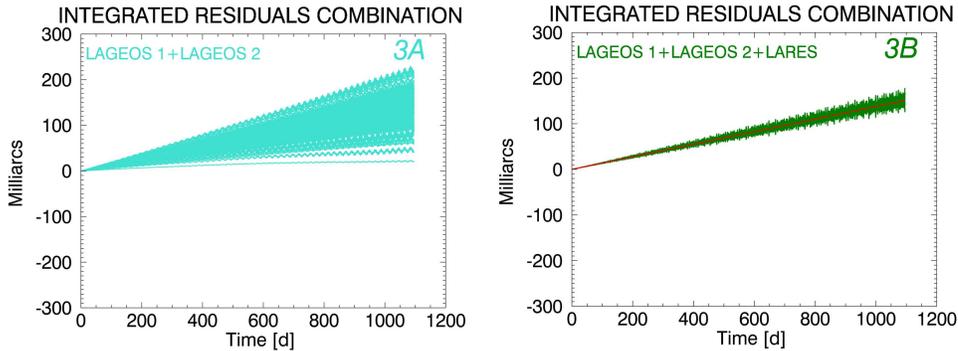}
\caption{Figure 3A, using a scale magnified by about 33.3 times with respect to Figure 2, shows the simulated cumulative residuals of the combination of the nodal longitudes of LAGEOS and LAGEOS 2, for each of the 100 simulations. That combination allows to eliminate the uncertainty in the Earth's quadrupole moment $J_2$. Figure 3B is the same of figure 2D but using a scale magnified by about 33.3 times (the same scale of figure 3A), i.e., it shows the simulated cumulative residuals of the combination of the nodal longitudes of LAGEOS, LAGEOS 2 and LARES that allows to eliminate the uncertainties $\delta J_2$ and $\delta J_4$. Also shown in 3B, in red, is the theoretical prediction of General Relativity. The reduction of the spread between the 100 simulations of Figure 3B with respect to Figure 3B clearly displays the improvement in the uncertainty of the measurement of frame-dragging using the LARES satellite.}
\end{center}
\end{figure}

\section{Conclusions}

The three observables provided by the three nodes of the LARES, LAGEOS and LAGEOS 2 satellites,
together with the Earth gravitational field determinations from the GRACE space mission, will allow to
improve significantly the previous measurements of the phenomenon of frame-dragging predicted by General Relativity, 
by eliminating the uncertainties in the value of the first two even zonal harmonics of the Earth potential, 
$J_2$ and $J_4$.
The LARES space experiment will also allow other tests of fundamental physics, such as the tests of String 
Theories of the type of Chern-Simons gravity described in section 1.1.

A number of previous detailed and extensive error analyses have confirmed the potentiality of the LARES
experiment to achieve a measurement of frame-dragging with an uncertainty of a few percent only.
However, to further test the previous extensive error
analyses, we have designed and performed 100 Monte Carlo simulations of the LARES experiment. In our
Monte Carlo analysis we have simulated the orbits of the LARES, LAGEOS and LAGEOS 2 satellites by
randomly generating the values of the GM (mass) of Earth, of its five largest even zonal harmonics,
$J_2, \, J_4, \, J_6, \, J_8$ and $J_{10}$,
of the secular rate of change of the two largest even zonal harmonics, $\dot {J_2}$ and $\dot {J_4}$, and of
the solar radiation coefficients of LARES, LAGEOS and LAGEOS 2. These parameters are identified as
the main source of bias in the measurement
of frame-dragging using LARES, LAGEOS, LAGEOS 2 and the GRACE-derived Earth gravitational field.

Future Monte Carlo simulations, in addition to the
secular rate of change of the first two even zonal harmonics considered here, will include other time variable
gravity effects, such as tides and secular changes
of higher even zonal harmonics and, in addition to direct solar radiation pressure and to the uncertainties in
the solar radiation coefficients of the three satellites considered here, will include other non-gravitational effects,
such as Earth's albedo and thermal drag. However, since these error sources are
constrained at the one percent level only, we do not expect any significant change in the final result of the
Monte Carlo simulations of the present paper, that is, an uncertainty, due to {\it systematic errors}, of about 1\% in
the measurement of frame-dragging using the LARES, LAGEOS and LAGEOS 2 satellites. The
standard deviation of the frame-dragging effect measured in the 100 simulations, representing the systematic 
errors in the measurement of frame-dragging, was in fact 1.4\% only of the total frame-dragging effect. 
Furthermore, on the basis of the preliminary, rough analyses
of the real orbits of the LARES, LAGEOS and LAGEOS 2 satellites, the {\it statistical error} in the measurement of 
frame-dragging was estimated to be about 1\% of the total frame-dragging effect.

\section{Acknowledgements}

The work is performed under the ASI contract n. I/034/12/0 and ESA contract n. 4000103504/2011/NL/WE.
The authors acknowledge the Italian Space Agency for its support to the LARES mission and the International Laser Ranging Service for tracking and data distribution of the LARES satellite. E.C. Pavlis acknowledges the support of NASA grant NNX09AU86G. We thank John Ries and the anonymous referees for useful suggestions.



\section*{References}
\begin{thebibliography}{10}


\bibitem{book1}  Ciufolini I,  Paolozzi A,  Pavlis E C,  Ries J,  Koenig R, Matzner R, Sindoni G and Neumayer H 2011 Testing Gravitational Physics with
Satellite Laser Ranging  \textit{The European Physical Journal Plus}  \textbf{126} 72

\bibitem{book2}  Ciufolini I, Paolozzi A,  Paris C Overview of the LARES Mission: orbit, error analysis and technological aspects 2012 \textit{Journal of Physics}, \textbf{354} 012002.

\bibitem{book3}  Williams, J G, Turyshev S G and  Murphy T W Jr 2004 Improving LLR tests of
gravitational theory  \textit{Int. J. Mod. Phys. D} \textbf{13} 567--82

\bibitem{book4} Cohen S C and Dunn P J 1985 LAGEOS scientific results  \textit{J. Geophys. Res. B} \textbf{90} 9215--9438.

\bibitem{book5} Noomen R, Klosko S, Noll C and Pearlman M (eds) Toward Millimeter Accuracy.
NASA CP 2003-212248 \textit{Proc. 13th Int. Laser Ranging Workshop} (NASA Goddard,
Greenbelt, Maryland, )

\bibitem{book6}  Ciufolini I,  Paolozzi A, Pavlis E C, Ries J, Gurzadyan V, Koenig R,  Matzner R,  Penrose R and  Sindoni G 2012 Testing General Relativity and gravitational physics using the LARES satellite \textit{The European Physical Journal Plus} \textbf{127}, 127

\bibitem{book7} Reigber C et al 2002 GRACE orbit and gravity field recovery at GFZ Potsdam-first experiences and perspectives \textit{Eos (Fall Meet. Suppl.)} \textbf{83} 47, abstr. G12B-03

\bibitem{book8} Tapley B D 2002 The GRACE mission: status and performance assessment \textit{Eos (Fall Meet. Suppl.)} \textbf{83} (47), abstr. G12B-01 .

\bibitem{book9}  Einstein A, Letter to Ernst Mach. Zurich, 25 June 1913, in ref. [12], p. 544

\bibitem{book10} Lense J, Thirring H, \"Uber den Einfluss der Eigenrotation der Zentralkorper auf die Bewegung der Planeten und Monde nach der Einsteinschen Gravitationstheorie 1918 \textit{Phys. Z.} \textbf{19}, 156-163.

\bibitem{book11} Ciufolini I 2007 Dragging of Inertial Frames \textit{Nature (Review)}, \textbf{449}, 41-47.

\bibitem{book12} Misner C W,  Thorne K S and  Wheeler J A 1973 \textit{ “Gravitation”}, Freeman eds.

\bibitem{book13} Weinberg S 1972 Gravitation and Cosmology: Principles and Applications of the General Theory of Relativity (Wiley, New York).

\bibitem{book14} Ciufolini I and Wheeler J 1995 Gravitation and Inertia \textit{Princeton University Press}

\bibitem{book15} Thorne K S, Price R H and  Macdonald D A 1986 The Membrane Paradigm \textit{Yale University Press} NewHaven.

\bibitem{book16} Ciufolini I and Ricci F 2002 Time delay due to spin and gravitational lensing \textit{Class. and Quantum Grav.} \textbf{19} 3863.

\bibitem{book17} Ciufolini I and Ricci F 2002 Time delay due to spin inside a rotating shell \textit{Class. and Quantum Grav.} \textbf{19} 3875.

\bibitem{book18} Pugh G E 1959 Proposal for a Satellite Test of the Coriolis Prediction of General Relativity \textit{Weapons Systems Evaluation Group Research Memorandum}  \textbf{11} (The Pentagon, Washington).

\bibitem{book19} Schiff L I 1960 Possible new test of General Relativity Theory \textit{Phys. Rev. Lett} \textbf{4} 215-217

\bibitem{book20} Ciufolini I and  Pavlis E C 2004 A Confirmation of the General Relativistic Prediction of the Lense-Thirring Effect \textit{Nature}, \textbf{431}, 958-960

\bibitem{book21} Ciufolini I,  Pavlis E C,  Ries J C,  Koenig R,   Paolozzi A, Sindoni G and  Neumayer H 2010 Test of Gravitomagnetism with the LAGEOS and GRACE Satellites \textit{General Relativity and John Archibald Wheeler}, ed  I Ciufolini and R Matzner  371-434 (Springer)

\bibitem{book22}  Everitt C W  F et al 2011 Gravity Probe B: Final Results of a Space Experiment to Test General Relativity \textit{ Phys. Rev. Lett}. \textbf{106}, 221101

\bibitem{book23}  Will C 2011 Finally, results from Gravity Probe-B. Comments: A Viewpoint article \textit{Physics 4} \textbf{43}

\bibitem{book24}   Smith T,  Erickcek A ,  Caldwell R R  and  Kamionkowski M 2008 The effects of Chern-Simons gravity on bodies orbiting the earth \textit{Phys. Rev. D} \textbf{77}, 024015

\bibitem{book25}  Ciufolini I 1986 Measurement of the Lense-Thirring Effect on Lageos and Another High Altitude Laser Ranged Artificial Satellite \textit{Physical Review Letters} \textbf{56}, 278-281

\bibitem{book26} Ciufolini I 1989 A Comprehensive Introduction to the Lageos Gravitomagnetic Experiment, from the Importance of the Gravitomagnetic Field in Physics to a Preliminary Error Budget \textit{Int. Journ. of Phys. A}  \textbf{4}, 3083-3145

\bibitem{book27} Ciufolini I 1996 On a new method to measure the gravitomagnetic field using two orbiting satellites \textit{Nuovo Cimento. A}  \textbf{109}, 1709-1720

\bibitem{book28}  Ciufolini I , Paolozzi A,  Pavlis E C, Ries J C,  Koenig R Matzner R and Sindoni G 2010 The LARES Relativity Experiment \textit{General Relativity and John Archibald Wheeler}, I. Ciufolini and R. Matzner eds. Springer, 467-492

\bibitem{book29} Tapley B, Ries J C, Eanes  R J, and Watkins M M 1989  \textit{ NASA-ASI Study on LAGEOS III} , CSR-UT publication n. CSR-89-3, Austin, Texas  and Ciufolini, I, et al 1989 \textit{ASI-NASA Study on LAGEOS III}, CNR, Rome, Italy

\bibitem{book30} Rubincam D P 1990 Drag on LAGEOS satellite \textit{J. Geophys. Res.} \textbf{ 95 ( B11 )}, 4881-4886

\bibitem{book31} Kaula W M  1966  \textit{Theory of Satellite Geodesy}  Blaisdell

\bibitem{book32} Petit G and Luzum B (eds.) 2010 IERS Conventions IERS Technical Note 36,  Frankfurt am Main \textit{Verlag des Bundesamts für Kartographie und Geodaesie} 179 pp., ISBN 3-89888-989-6.

\bibitem{book33} Lucchesi D M 2002 Reassessment of the error modelling of non—gravitational perturbations on LAGEOS 2 and their impact in the Lense—Thirring determination \textit{ Planet. Space Sci. Part II,} \textbf{50} 106

\bibitem{book34} Bruinsma S L, Marty J C, Balmino G, Biancale R, Foerste C, Abrikosov O and Neumayer H 2010 GOCE Gravity Field Recovery by Means of the Direct Numerical Method \textit{ESA Living Planet Symposium 2010}, Bergen, June 27 - July 2 2010, Bergen, Norway.

\bibitem{book35} Zhu S, Reigber Ch, and Koenig R 2004 Integrated Adjustment of CHAMP, GRACE, and GPS Data \textit{Journal of Geodesy} \textbf{78}, 103-108




\end{thebibliography}
\end{document}